\documentclass{INTERSPEECH2023}
\usepackage{microtype}
\usepackage{graphicx}
\usepackage{subfigure}
\usepackage{booktabs} 
\usepackage{url}
\usepackage{comment}
\usepackage{caption}
\usepackage{makecell}
\usepackage{amsmath}
\usepackage{amssymb}
\usepackage{mathtools}
\usepackage{amsthm}
\usepackage{multirow}
\usepackage{multicol}
\mathchardef\mhyphen="2D

\interspeechcameraready

\title{Laughter Synthesis using Pseudo Phonetic Tokens with a Large-scale In-the-wild Laughter Corpus}
\name{Detai Xin, Shinnosuke Takamichi, Ai Morimatsu, Hiroshi Saruwatari}
\address{
  Graduate School of Information Science and Technology, The University of Tokyo, Japan}
\email{\{detai\_xin, shinnosuke\_takamichi\}@ipc.i.u-tokyo.ac.jp}

\begin{document}

\maketitle

\begin{abstract}
\vspace{-2mm}
We present a large-scale in-the-wild Japanese laughter corpus and a laughter synthesis method.
Previous work on laughter synthesis lacks not only data but also proper ways to represent laughter.
To solve these problems, we first propose an in-the-wild corpus comprising $3.5$ hours of laughter, which is to our best knowledge the largest laughter corpus designed for laughter synthesis.
We then propose pseudo phonetic tokens (PPTs) to represent laughter by a sequence of discrete tokens, which are obtained by training a clustering model on features extracted from laughter by a pretrained self-supervised model.
Laughter can then be synthesized by feeding PPTs into a text-to-speech system.
We further show PPTs can be used to train a language model for unconditional laughter generation.
Results of comprehensive subjective and objective evaluations demonstrate that the proposed method significantly outperforms a baseline method, and can generate natural laughter unconditionally.
\end{abstract}
\vspace{-5mm}
\section{Introduction}
\vspace{-2mm}
Human speech contains not only verbal but also nonverbal expressions like laughter, sobbing, and scream, etc.~\cite{scherer1994affect, trouvain2012comparing}, which can effectively convey internal affects~\cite{hall2009psychosocial, scherer2011assessing} of speakers in various languages and cultures~\cite{sauter2010cross}.
Although recent advances in speech synthesis are able to synthesize natural verbal speech that is indistinguishable from human speech~\cite{ren2019fastspeech,kim2020glow, ren2020fastspeech, kong2020hifi}, the progress in synthesizing nonverbal expressions is limited due to the lack of both data and technologies.
In this work, we focus on a typical but important task in nonverbal-expression synthesis: laughter synthesis.
Being able to synthesize laughter can intuitively improve the expressiveness and authenticity of a speech synthesis system.
Such systems can be applied, for example, in virtual agents to smooth communication with users~\cite{el2015speech}.

In most previous work, how to define a proper representation of laughter seems to be a core problem for laughter synthesis.
Haddad et al.~\cite{el2015speech} and Nagata et al.~\cite{nagata2018defining} manually transcribe laughter into phonemes, and use HMM-based models to synthesize laughter.
Such methods are prohibitively costly for modern data-driven methods based on deep neural networks (DNNs).
Mori et al.~\cite{mori2019conversational} propose to input power contours of laughter into WaveNet~\cite{vanwavenet} to synthesize laughter.
However, it cannot control the phonetic content of laughter.
Besides, more abstractive presentations like latent variables ~\cite{mansouri2020laughter, afsar2021generating} and emotion labels~\cite{matsumoto2020controlling} are also used.
Most recently, Luong et al.~\cite{luong2021laughnet} propose an abstractive representation of laughter called silhouette that is obtained by framing the original waveforms with min and max pooling.
Laughter is synthesized by HiFi-GAN~\cite{kong2020hifi} from the silhouette.
All aforementioned abstractive representations can avoid troublesome human annotations, but on the other hand, they have low controllability in the synthesis process compared to phonemes that can be easily manipulated.

Another critical problem for laughter synthesis is a lack of data.
The number of open-sourced laughter corpus that is suitable for laughter synthesis is quite limited.
Therefore, researchers usually have to use a verbal corpus mixed with a small number of laughter~\cite{arimoto2012naturalistic} or even buy a commercial corpus with limited size~\cite{luong2021laughnet}, which further impedes more works in this task.

In this paper, we present a method for laughter synthesis using pseudo phonetic tokens on a large-scale in-the-wild laughter corpus.
To solve the problem of the lack of data, we first propose a new Japanese laughter corpus collected from the Internet.
In the proposed method, firstly a clustering model based on k-means~\cite{macqueen1967classification} is trained on features extracted from the laughter utterances by a self-supervised learning (SSL) model called HuBERT~\cite{hsu2021hubert}.
The clustering model is then used to transcribe each utterance into a sequence of discrete tokens containing the phonetic information of the original laughter, which we call pseudo phonetic tokens (PPTs).
A Text-to-speech (TTS) model is then trained by regarding PPTs as text inputs to synthesize laughter.
The proposed phonetic token as a representation of laughter not only avoids human annotation but also has higher controllability than the previous abstractive representations aforementioned.
Furthermore, we show it is possible to train a token language model (tLM) on the PPTs to enable unconditional laughter synthesis.
We conduct comprehensive objective and subjective experiments on the proposed corpus.
Experimental results demonstrate that: (1) the proposed method significantly outperforms a baseline method that uses phonemes to represent laughter; (2) the proposed method can generate natural laughter unconditionally with the assistance of tLM.
The contributions of this work are summarized as follows:
\vspace{-1.5mm}
\begin{itemize} \itemsep -1mm 
    \item We propose a large-scale in-the-wild Japanese laughter corpus. This corpus is, to our best knowledge, currently the largest laughter corpus that is suitable for laughter synthesis.
    \item We propose a method for laughter synthesis using pseudo phonetic tokens as the representation of laughter.
    \item We propose to train a token language model to generate PPTs and synthesize laughter unconditionally.
    \item We conduct comprehensive objective and subjective experiments to demonstrate the proposed method can synthesize natural laughter that is significantly better than a baseline method.
\end{itemize}
\vspace{-1.5mm}
We publicate the proposed corpus\footnote{\url{sites.google.com/site/shinnosuketakamichi/research-topics/laughter_corpus}} and the code implementation\footnote{\url{github.com/Aria-K-Alethia/laughter-synthesis}} of the proposed method.
\vspace{-3mm}
\section{Laughter data collection}
\vspace{-2mm}
\begin{figure}[t]
\begin{center}
\centerline{
\includegraphics[width=\columnwidth]{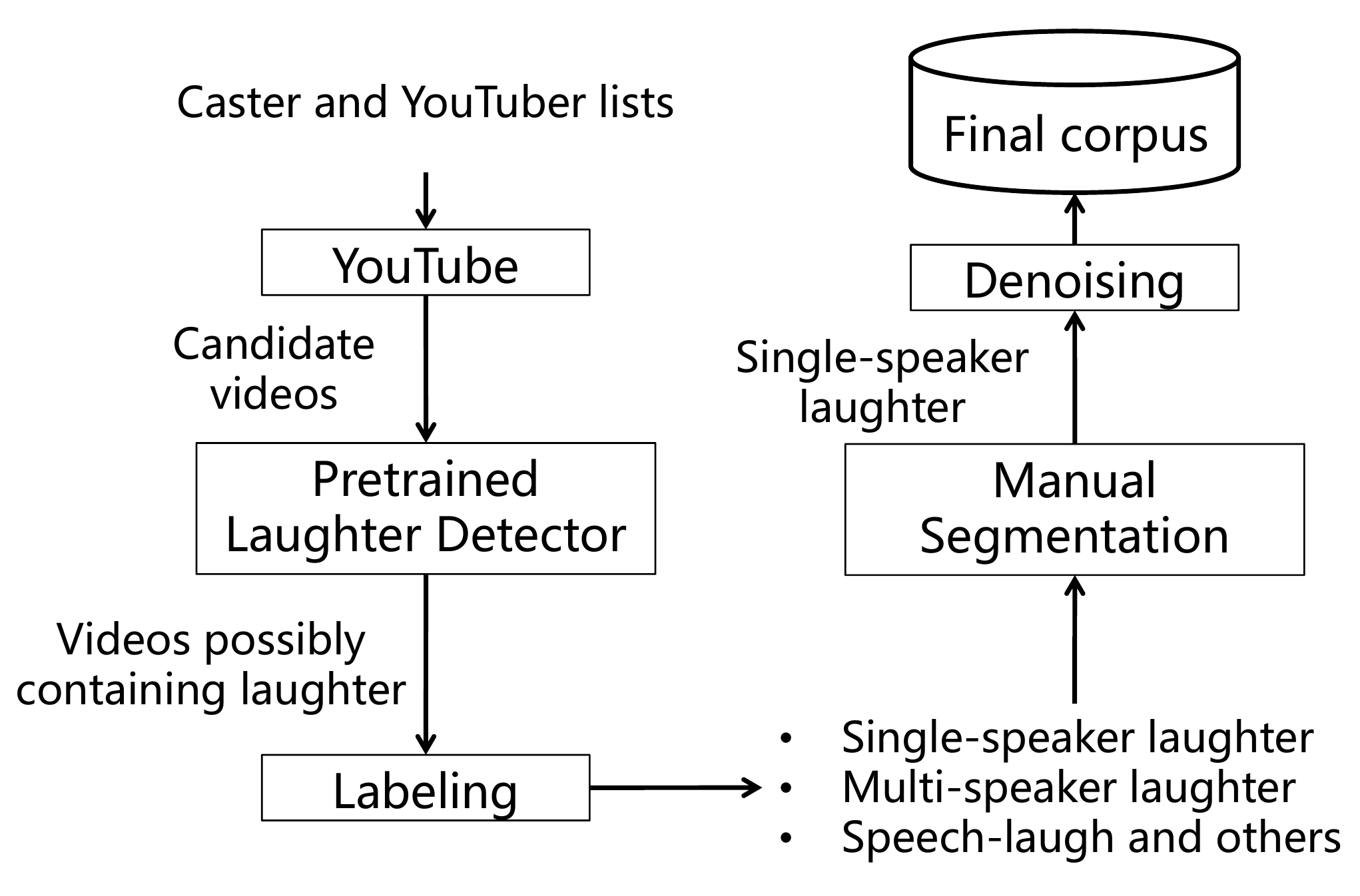}
}
\vspace{-5mm}
\caption{Data-collection process for the proposed corpus.}
\label{figure:data_collection}
\vspace{-12mm}
\end{center}
\end{figure}

We aim to collect large-scale in-the-wild laughter utterances that are suitable for laughter synthesis.
The general data-collection process is illustrated in Figure~\ref{figure:data_collection}.
We first use several lists of casters and YouTubers obtained from Wikipedia~\footnote{For example, \url{en.wikipedia.org/wiki/List_of_YouTubers}} to crawl candidate videos during June 2022 by searching the names in the list on YouTube, which results in about $10$k videos.
Using the lists ensures we only collect human speech instead of others like animal sounds.
Second, we use an open-sourced pretrained laughter detection model~\footnote{\url{github.com/jrgillick/laughter-detection}} to discover videos that possibly contain laughter, which results in about $1500$ videos.
However, we find that many of the detected videos include multi-speaker laughter or speech-laugh which are not suitable for synthesis.
Therefore, we further conduct a listening test with crowd-sourcing to label the detected videos.
Specifically, we request about $1500$ workers to label the videos with three categories: (1) single-speaker laughter; (2) multi-speaker laughter; (3) others including speech laugh.

After labeling, we manually segment laughter utterances from those videos that have at least one ``single-speaker laughter" label.
Note that, many videos contain background noises, and we only select those with non-speech noise to simplify the denoising step.
Besides, we discard non-Japanese videos in this step.
Finally, to reduce noises in the utterances, we use a source separation model called Demucs\footnote{\url{github.com/facebookresearch/demucs}}, which is a powerful source separation model based on DNNs, to extract the vocals from the videos.
Specifically, we use the pretrained Demucs v3 (``\texttt{\detokenize{hdemucs_mmi}}") model~\cite{defossez2021hybrid} since we find it is more stable than the latest v4 model.
The final corpus contains $7489$ utterances of single-speaker laughter from $470$ speakers.
The total duration of the corpus is about $3.5$ hours.

Arimoto et al. propose OGVC that has $1669$ laughter utterances mixed with verbal speech~\cite{arimoto2012naturalistic}.
The total duration of laughter in OGVC is about $0.5$ hours.
Cowen et al. propose H-VB, which is a multilingual nonverbal-expression corpus~\cite{Cowen2022HumeVB}
The total duration of H-VB is about $36$ hours, but unfortunately, it is a private corpus.
Some corpora like AudioSet~\cite{gemmeke2017audio} do have nonverbal expressions but are not suitable for laughter synthesis since they have multi-speaker laughter.
Therefore, to our best knowledge, the proposed corpus is currently the largest open-sourced single-speaker laughter corpus that is suitable for laughter synthesis.
\vspace{-4mm}
\section{Laughter synthesis using pseudo phonetic tokens}
\vspace{-2mm}
\begin{figure}[t]
  \centering
  \subfigure[Laughter representations]{
        \centering
        \includegraphics[width=0.5\linewidth]{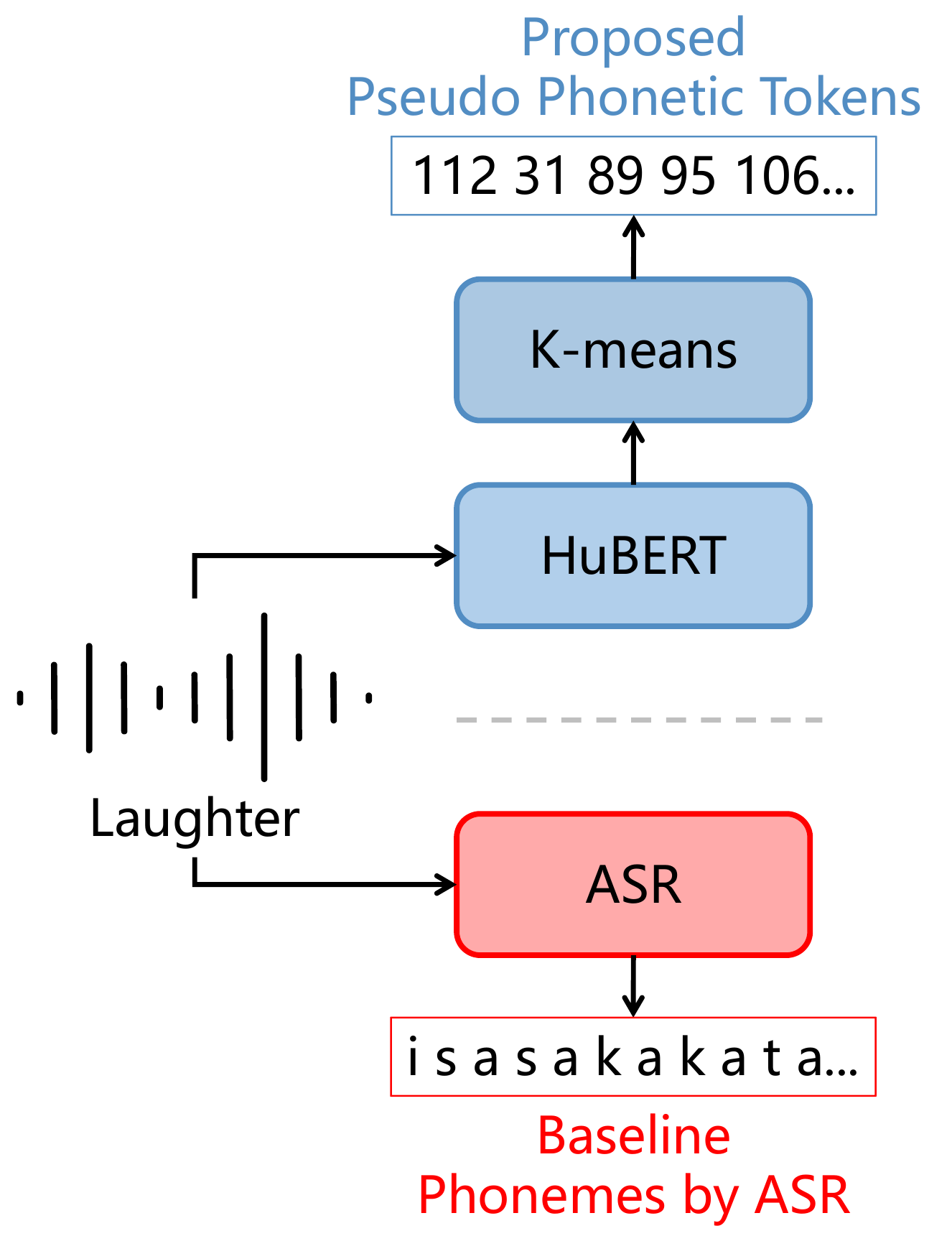}
  }
  \subfigure[TTS architecture]{
        \centering
        \includegraphics[width=0.5\linewidth]{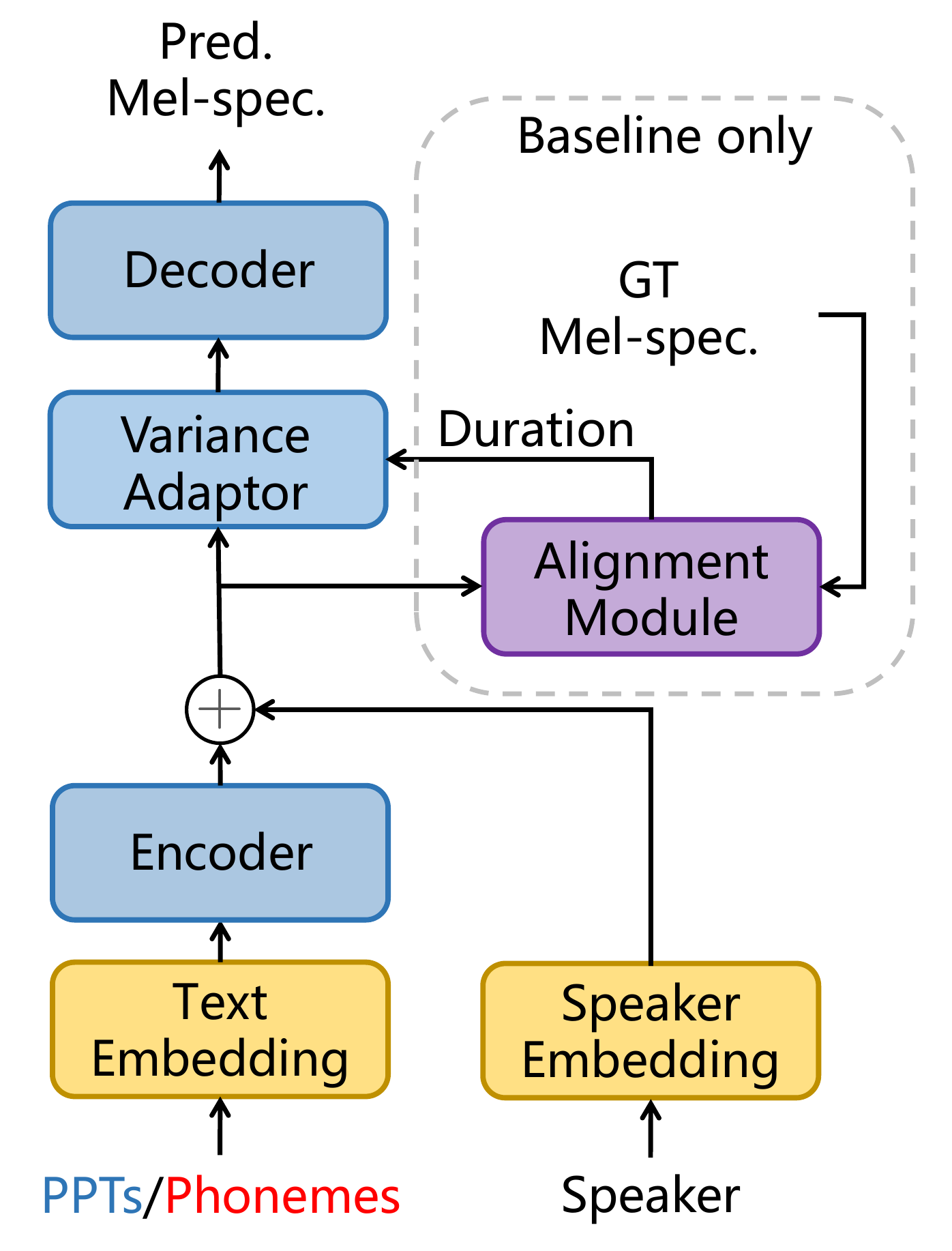}
  }
  \vspace{-3mm}
  \caption{Left: laughter representations used in this work; Right: architecture of the TTS model. For the baseline method an additional alignment module is used.}
  \label{fig:architecture}
  \vspace{-7mm}
\end{figure}
The general architecture of the proposed and baseline methods is illustrated in Figure~\ref{fig:architecture}.
The major difference between the two methods is that they use different methods to transcribe laughter (Figure~\ref{fig:architecture} (a)).
Both methods use a TTS model to synthesize mel-spectrograms from the obtained transcriptions (Figure~\ref{fig:architecture} (b)).
In this section, we first introduce the baseline method together with the TTS model.
Then, we describe each component of the proposed method separately.

\vspace{-3mm}
\subsection{The baseline method}
\vspace{-2mm}
As illustrated in the bottom half of Figure~\ref{fig:architecture} (a), we use a pretrained multilingual ASR model based on wav2vec 2.0~\cite{xu2021simple} to transcribe laughter into phoneme sequences.
We then adopt FastSpeech2~\cite{ren2020fastspeech} to synthesize mel-spectrograms from the phoneme sequences.
To enable the model to support multi-speaker synthesis, we incorporate a look-up speaker embedding table in the model.
Besides, the original FastSpeech2 relies on an external alignment tool to get the duration information for each phoneme, but it is difficult to find an off-the-shelf alignment tool for the standard International Phonetic Alphabet (IPA) used by the multilingual ASR model in the baseline method.
Therefore, we use an unsupervised alignment module inspired by Glow-TTS~\cite{kim2020glow} that can be jointly trained with the TTS model.
The alignment module receives the output of the phoneme encoder and the ground truth (GT) mel-spectrogram as input and outputs a probability distribution for each frame of the mel-spectrogram over the phoneme sequence, which is called the soft alignment matrix.
The duration information of each phoneme can then be retrieved by using a Viterbi-like algorithm that is called the monotonic alignment search in the original paper~\cite{kim2020glow} to binarize the distributions.
The binarized distribution is called the hard alignment matrix.
The module can be efficiently trained in an unsupervised manner using the connectionist temporal classification (CTC) loss~\cite{graves2006connectionist}.
In addition, we also use a binarization loss based on KL-divergence to minimize the distance between the outputted distributions and the binarized distributions~\cite{shih2021rad}.
Readers are recommended to refer to the original papers~\cite{kim2020glow, shih2021rad} for more details.
The final loss value of the baseline method is a summation of the FastSpeech2 loss and the loss of the alignment module.

\vspace{-3mm}
\subsection{Pseudo phonetic tokens}
\vspace{-2mm}
Transcriptions based on ASR have two problems: (1) since the ASR model is mostly trained on verbal speech, the predicted transcriptions may be imprecise for nonverbal laughter; (2) some laughter utterances in the proposed corpus are too short to transcribe.
Actually, the ASR model used in the baseline method has a failure rate of about $6.9\%$ in the experiments and outputs empty transcriptions for those utterances.

To solve the above problems, we propose pseudo phonetic tokens (PPTs) to represent laughter.
The proposed PPT is inspired by generative spoken language modeling~\cite{lakhotia2021generative, kreuk2021textless}, which originally uses SSL models to discretize speech to do TTS in a textless manner.
In this work, we further adapt this idea into nonverbal laughter.
As shown in the top half of Figure~\ref{fig:architecture} (a), the waveform is first fed into HuBERT~\cite{hsu2021hubert} to convert it into continuous sequential features.
Then, a k-means model~\cite{macqueen1967classification} is trained upon the features, which can be used to convert the continuous features into discrete tokens (cluster indices).
Although the continuous features encode rich information of the original waveforms including not only linguistic information but also non-linguistic information like speaker identities and prosody, the discretization process can factor out most non-linguistic information, as shown in a previous work~\cite{van2017neural}.
Hence, we call the discretized representations as pseudo phonetic tokens.
In this work, we only use HuBERT as the SSL model since several previous works have shown HuBERT is better than other SSL models like wav2vec 2.0 in representing speech as discrete tokens~\cite{lakhotia2021generative, polyak2021speech}.
Since HuBERT is trained with a self-supervised criterion without using transcriptions, we believe it is more appropriate to use it for nonverbal expressions than normal ASR models.
Also, the SSL model can transcribe all utterances in the proposed corpus with no failure.

The obtained PPTs are then fed into a TTS model to synthesize laughter.
The TTS model has all components used in the baseline method except for the alignment module.
This is because the running length of each PPT can be regarded as its duration.
For example, for a PPT sequence $[21, 21, 34, 21]$, its duration sequence is $[2, 1, 1]$.
Following original GSLM~\cite{lakhotia2021generative}, we remove sequential repetitions (the sequence in the above example becomes $[21, 34, 21]$ after removing) in all PPT sequences before inputting them to the phoneme encoder.

\vspace{-3mm}
\subsection{Token language model}
\vspace{-2mm}
PPT can be regarded as a symbolic representation of laughter.
Thus, it is possible to train a token language model (tLM) on the PPTs of the proposed corpus.
After training, one can generate laughter unconditionally by sampling from tLM.
Such unconditional generation is, however, intuitively difficult for the abstractive representations proposed in the previous work~\cite{mori2019conversational, mansouri2020laughter, afsar2021generating, matsumoto2020controlling, luong2021laughnet}, which demonstrates the proposed PPTs have higher controllability.
Actually, a similar idea is also proposed in the original GSLM~\cite{lakhotia2021generative}, which uses a token language model for verbal speech generation.
We argue that tLM is more suitable for modeling nonverbal expressions.
First, the patterns of nonverbal expressions are simpler than verbal speech, which makes it easy to train a language model on PPTs.
Second, verbal speech can be easily transcribed by ASR, but this is intrinsically difficult for nonverbal expressions like laughter, which makes tLM more necessary for nonverbal expressions.
Finally, PPTs factor out semantic information compared to linguistic tokens like words.
Such information is essential for understanding verbal speech, but not for nonverbal expressions.
\vspace{-3mm}
\section{Experiments}
\vspace{-3mm}
\subsection{Setup}
\vspace{-2mm}
We downsample all waveforms into 16~kHz.
Since the fps of HuBERT is $50$, we set hop length to $320$ to extract all acoustic features including pitch and mel-spectrograms.
The pitch information of each utterance is extracted with WORLD vocoder~\cite{morise2016world}.

We exclude utterances that are too long (over $20$ s) or cannot get pitch values by the WORLD vocoder, which results in $7290$ utterances.
We split these utterances into train/validation/test sets with $7110$/$90$/$90$ utterances, respectively.
The test set consists of $30$ speakers with $3$ utterances per speaker, which are randomly selected from the speakers who have at least $10$ utterances in the proposed corpus.

We use a pretrained multilingual wav2vec 2.0 model (XLSR)~\cite{xu2021simple} fine-tuned on CommonVoice\footnote{\url{huggingface.co/facebook/wav2vec2-xlsr-53-espeak-cv-ft}}~\cite{ardila2019common} as the multilingual ASR model used in the baseline method.
The resulting transcriptions have $87$ unique symbols in IPA.

We use the pretrained ``\texttt{\detokenize{hubert-base-ls960}}" model\footnote{\url{huggingface.co/facebook/hubert-base-ls960}} to extract the continuous sequential features used in the proposed method.
This model is trained on the $960$-hour LibriSpeech corpus~\cite{panayotov2015librispeech} with a $12$-layer transformer-based architecture~\cite{hsu2021hubert}.
For k-means clustering, we use the implementation of sklearn\footnote{\url{scikit-learn.org/stable/modules/generated/sklearn.cluster.MiniBatchKMeans.html}} to train the model.
We set the cluster number to $200$, which means that there are $200$ different PPTs used in the TTS model.
The batch size is set to $10000$.
We train several k-means models; most of them converge in about $250$ iterations.
After training, we convert all utterances into their PPT representations.

We use the same architecture of the original FastSpeech2~\cite{ren2020fastspeech}.
The dimension of the speaker embedding is set to $256$.
For the alignment module in the baseline method, we use exactly the same training strategy used in RAD-TTS~\cite{shih2021rad}.
Specifically, we start to binarize the soft alignment matrix after $6$k steps; we enable the binarization loss based on KL-divergence after $18$k steps; we add an alignment prior formulated by a beta-binomial distribution into the soft alignment matrix to accelerate the alignment learning.
For all TTS models, the batch size is set to $16$.
Adam~\cite{kingma2014adam} is used as the optimizer with a scheduled learning rate proposed in~\cite{vaswani2017attention}.
All models converge in about $200$k steps.

We use HiFi-GAN~\cite{kong2020hifi} as the vocoder to convert mel-spectrograms into time-domain waveforms.
As the hop length of the officially released pretrained models is not $320$, we train a new HiFi-GAN vocoder from scratch on a multi-speaker Japanese corpus~\cite{takamichi2019jvs}.
We use the official script\footnote{\url{github.com/jik876/hifi-gan}} to train the model.
The training takes about $1.5$ weeks on an NVIDIA V100 GPU card.

We use fairseq~\cite{ott2019fairseq} to train tLMs.
We use the ``\texttt{\detokenize{transformer_lm}}" architecture, which is based on a $6$-layer transformer~\cite{vaswani2017attention}.
Adam~\cite{kingma2014adam} is used as the optimizer with an initial learning rate of $5e\mhyphen4$.
The batch size is set to $16$.
All tLMs converge with about $30$ epochs.
After training, we generate $90$ sequences of PPTs unconditionally for each tLM.
The temperature is set to $0.7$.
These sequences are then inputted into the TTS model to synthesize laughter with the same speaker setting of the test set.

\vspace{-3mm}
\subsection{Objective metrics}
\vspace{-2mm}
We use several objective metrics computed on the test set or generated sequences of PPTs of laughter to evaluate the TTS models and tLMs:
\begin{itemize} \itemsep -1mm 
    \item \textbf{Mel-cepstral distortion (MCD)} computed with dynamic time warping (DTW).
    \item \textbf{F0 root mean square error (F0-RMSE)} computed with DTW.
    \item \textbf{Perplexity (PPL)} defined as the normalized inverse probability on the test set of the tLM.
    \item \textbf{Self-BLEU~\cite{zhu2018texygen}} defined as the average value of the $n$-gram ($4$-gram in this work) BLEU scores~\cite{papineni2002bleu} between one generated sentence and the rest generated sentences for all generated sentences.
\end{itemize}
Here MCD and F0-RMSE reflect the quality of the synthesized laughter; PPL and Self-BLEU reflect the performance of the tLM and the diversity of the generated sentences, respectively.
In particular, since each tLM has a unique set of PPTs, in this work we propose to use a normalized version of Self-BLEU that is defined as the ratio of the Self-BLEU of the generated sentences to the Self-BLEU of the test set: $\overline{\mathrm{Self \mhyphen BLEU}}=\mathrm{Self \mhyphen BLEU}/\mathrm{Self \mhyphen BLEU}_{\mathrm{gt}}$.
This metric has a value between $[0, 1]$, and can reflect how diverse the generated sentences are compared to the GT sentences.

\begin{figure}[t]
\begin{center}
\centerline{
\includegraphics[width=\columnwidth]{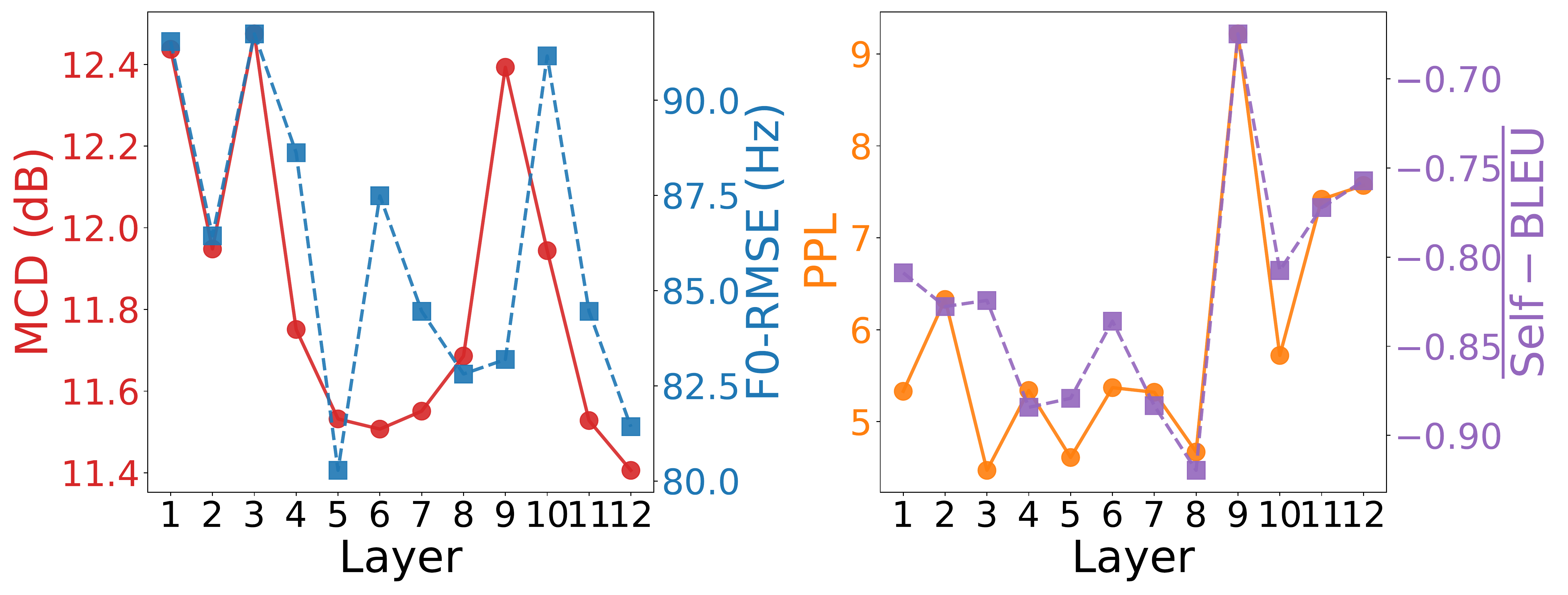}
}
\vspace{-3mm}
\caption{Objective performance of the proposed method using the output of different layers of HuBERT as the feature for PPTs. Negative $\overline{\mathrm{Self \mhyphen BLEU}}$ scores are shown for the ease of comparison.}
\label{figure:obj_layer}
\vspace{-8mm}
\end{center}
\end{figure}

\vspace{-3mm}
\subsection{Laughter synthesis}
\vspace{-2mm}
\subsubsection{Layer selection of HuBERT}
\vspace{-2mm}
As the output of each layer of HuBERT is possible to be used as the features for PPTs, we train $12$ proposed models and compute the objective metrics to select the best layer.
The result is illustrated in Figure~\ref{figure:obj_layer}.
Hereafter we use L$\{1,2,...,12\}$ to denote the proposed method using the corresponding layer of HuBERT for simplicity.
It can be seen that the performances of the TTS models (left) and the tLMs (right) are not consistent.
Specifically, L$12$ has the best performance among the TTS models, but L$8$ has the best performance among the tLMs.
Besides, L$5$ has good performance in all metrics.
Therefore, we use L$5$, L$8$, and L$12$ in the following evaluations.
We also tried the proposed method with a fewer or larger cluster number but found no improvements in the preliminary experiments.

\begin{table}[t]
    \centering
    \caption{Performance of all methods in the evaluations of laughter synthesis. \textbf{Bold} indicates the best score with $p < 1e\mhyphen5$ comparing to Baseline.}
    \vspace{-3mm}
    \footnotesize
    \begin{tabular}{c|cc|cc}
    \hline
    Model & MCD($\downarrow$) & F0-RMSE($\downarrow$) & MOS($\uparrow$) & SMOS($\uparrow$)\\
    \hline\hline
    GT       &   -      &  -     &   $3.73$  &  - \\
    HiFi-GAN & $6.68$  & $53.70$ & $3.31$  & $4.74$ \\
    \hline 
    Baseline & $16.59$  & $117.65$ & $1.25$  & $1.20$\\
    Baseline GT & $10.74$  & $85.69$ & -    &   -\\
    Proposed-L$5$ &  $11.53$  & $\mathbf{80.28}$ &  $\mathbf{3.00}$  & $3.07$\\
    Proposed-L$8$ & $11.69$  & $82.81$ & $2.98$  & $3.17$\\
    Proposed-L$12$ & $\mathbf{11.41}$  & $81.43$ & $2.96$  & $\mathbf{3.22}$ \\
    \hline
    \end{tabular}
    \label{tab:test}
    \vspace{-5mm}
\end{table}
\vspace{-3mm}
\subsubsection{Comparison to the baseline}
\vspace{-2mm}
Next, we compare the proposed method to the baseline method.
In addition to the objective metrics, we also use subjective mean opinion score (MOS) and similarity MOS (SMOS) to evaluate the naturalness and similarity of the synthesized laughter, respectively.
We conduct a standard $5$-scale MOS test on a Japanese crowd-sourcing platform\footnote{\url{https://www.lancers.jp/}} to compute the MOS of the synthesized $90$ utterances of each model, including the GT utterances and the utterances synthesized by HiFi-GAN from the GT mel-spectrograms.
$54$ listeners join in this test; each evaluates $36$ utterances of which the first $6$ are dummy samples used to enable the workers to get familiar with the task.
The answers of the dummy samples are not counted in the final result.
As a result, each utterance has $3$ answers.
The SMOS test is conducted in a similar setting with $45$ listeners, but the GT utterances are excluded.

All results are shown in Table~\ref{tab:test}.
First, the baseline method has poor performance in both the objective and subjective evaluations.
To verify if this is because the model fails to learn from the inputted phonemes, we further use GT acoustic features (pitch and energy) to synthesize the test utterances.
The corresponding model is denoted as ``Baseline GT" in Table~\ref{tab:test}.
It can be seen that the performance becomes comparable to the proposed method, which implies that the laughter representation makes the performance of the baseline method bad.
Second, it can be seen that the $3$ proposed models have significantly better performance than the baseline method in all metrics, which demonstrates the effectiveness of the proposed method using PPTs as the representation for laughter.
Finally, we observe that L$5$ has the best naturalness and L$12$ has the best speaker similarity, which is consistent with the results in objective metrics as both of the two models have better objective performance than L$8$.

\begin{table}[t]
    \centering
    \caption{Subjective performance of the proposed method in the evaluation of unconditional laughter generation. \textbf{Bold} indicates the best score with $p < 1e\mhyphen5$.}
    \vspace{-3mm}
    \footnotesize
    \begin{tabular}{c|cc}
    \hline
    Model & MOS($\uparrow$) & SMOS($\uparrow$)\\
    \hline\hline
    Proposed-L$5$  &  $\mathbf{3.11}$ & $\mathbf{2.65}$      \\
    Proposed-L$8$  & $2.80$ &  $2.59$      \\
    Proposed-L$12$ & $3.06$ &  $2.59$      \\
    \hline
    \end{tabular}
    \label{tab:subjective_ulm}
    \vspace{-5mm}
\end{table}
\vspace{-3mm}
\subsection{Unconditional laughter generation}
\vspace{-2mm}
Finally we evaluate the performance of unconditional laughter generation with a MOS test and a SMOS test.
Given the poor performance of the baseline method shown in the previous section, we only use the three proposed models in this evaluation.
$27$ listeners join in the MOS test; each evaluates $33$ utterances of which the first $3$ are dummy samples.
As a result, each utterance has $3$ answers.
The SMOS test is conducted using exactly the same setting of the MOS test.

The result is shown in Table~\ref{tab:subjective_ulm}.
It can be seen that generally L$5$ $>$ L$12$ $>$ L$8$.
This is quite different from the performance of tLMs shown in the right side of Figure~\ref{figure:obj_layer}, in which L$8$ $>$ L$5$ $>$ L$12$.
We suppose this is because the quality of the synthesized laughter is mainly determined by the performance of the TTS models.
However, it should be pointed out that MOS and SMOS cannot evaluate the diversity of the synthesized laughter subjectively.
We leave this as future work.
Combining this result with the result shown in Table~\ref{tab:test}, we conclude that layer $5$ is the best layer of HuBERT for PPTs used in laughter synthesis.
\vspace{-4mm}
\section{Conclusions}
\vspace{-3mm}
This paper first presented a large-scale in-the-wild Japanese laughter corpus designed for laughter synthesis, which is to our best knowledge the largest laughter corpus designed for laughter synthesis.
This paper then presented a laughter synthesis method using PPTs to represent laughter extracted by an SSL model and a clustering model.
Experimental results demonstrate: (1) the proposed method can synthesize natural laughter that is significantly better than a baseline method; (2) the proposed method can generate natural laughter unconditionally by training a token language model on PPTs.

{\normalsize
\textbf{Acknowledgements:} 
This work was supported by JST SPRING, Grant Number JPMJSP2108, JSPS KAKENHI, Grant Number JP23KJ0828, and JST FOREST JPMJFR226V. 
}

\bibliographystyle{IEEEtran}
\bibliography{mybib}

\end{document}